\documentclass[iop]{emulateapj}  

\usepackage{amsmath}
\newcommand\kms{\ifmmode{\rm km\thinspace s^{-1}}\else km\thinspace s$^{-1}$\fi}
\newcommand\vstar{HD\thinspace 28363}
\newcommand\hip{{\it Hipparcos\/}}
\newcommand\gaia{{\it Gaia\/}}

\shortauthors{Torres et al.}
\shorttitle{\vstar}

\begin{document} 

\submitted{Accepted for publication in The Astrophysical Journal}

\title{Dynamical masses for the triple system \vstar\ in the Hyades cluster}

\author{
Guillermo Torres,
Robert P.\ Stefanik, and
David W.\ Latham
}

\affiliation{Center for Astrophysics $\vert$ Harvard \&
  Smithsonian, 60 Garden St., Cambridge, MA 02138, USA;
  gtorres@cfa.harvard.edu}

\begin{abstract}

The star \vstar\ in the Hyades cluster has been known for over a
century as a visual binary with a period of 40 yr. The secondary is, in
turn, a single-lined spectroscopic binary with a 21-day period. Here we
report extensive spectroscopic monitoring of this hierarchical triple
system that reveals the spectral lines of the third star for the first
time.  Combined with astrometric information, this makes it possible
to determine the dynamical masses of all three stars. Only six other
binaries in the Hyades have had their individual component masses
determined dynamically.  We infer the properties of the system by
combining our radial velocity measurements with visual observations,
lunar occultation measurements, and with proper motions from the
\hip\ and \gaia\ missions that provide a constraint on the astrometric
acceleration. We derive a mass of $1.341^{+0.026}_{-0.024}~M_{\sun}$
for the visual primary, and $1.210 \pm 0.021$ and $0.781 \pm
0.014~M_{\sun}$ for the other two stars.  These measurements along
with those for the other six systems establish an empirical
mass-luminosity relation in the Hyades that is in broad agreement with
current models of stellar evolution for the known age and chemical
composition of the cluster.

\end{abstract}

%\keywords{
%binaries: visual;
%stars: fundamental parameters;
%stars: individual (\vstar);
%}

%%%%%%%%%%%%%%%%%%%%%%%%%%%%%%%%%%%%%%%%%%%%%%%%%%%%%%%%%%%%%%%%%%
\section{Introduction}
\label{sec:introduction}
%%%%%%%%%%%%%%%%%%%%%%%%%%%%%%%%%%%%%%%%%%%%%%%%%%%%%%%%%%%%%%%%%%

Relatively few binary or multiple systems in the nearby Hyades cluster
have had the dynamical masses determined for their individual
components. These are the essential ingredients for establishing the
empirical mass-luminosity relation (MLR) in the cluster, which serves
as a valuable test of current models of stellar evolution.  Recent
work by \cite{Torres:2019} has added one more system (80\thinspace
Tau) to the five classical binaries that have been used for this
purpose in the Hyades. They are V818\thinspace Tau
\citep{McClure:1982, Schiller:1987}, $\theta^2$\thinspace Tau
\citep{Peterson:1993, Tomkin:1995}, 51\thinspace Tau
\citep{Torres:1997a}, 70\thinspace Tau \citep{Torres:1997b}, and
$\theta^1$\thinspace Tau \citep{Torres:1997c}.

The formal precision of these determinations is typically between 5\%
and 15\%, with the exception of the 5.6-day eclipsing binary
V818\thinspace Tau, which has its masses measured to better than 1\%.
Studies of the Hyades MLR in addition to those cited above include the
papers by \cite{Lastennet:1999} and by \cite{Lebreton:2001}, who also
made a determination of the helium abundance of the cluster. More
recent work on Hyades binaries has mostly dealt with improvements
in the mass estimates of previously known systems.

In this paper, we present dynamical mass determinations for a new
system, \vstar\ (HIP\thinspace 20916, WDS\thinspace J04290+1610AB,
HU\thinspace 1080, ADS\thinspace 3248\thinspace AB, vB\thinspace 75;
$V = 5.56$), a visual binary with a 40-yr period discovered in 1904 by
\cite{Hussey:1905}. The secondary is itself a spectroscopic binary
with a period of about 21 days \citep{Stefanik:1992, Mermilliod:1994,
  Smekhov:1995}, making the system a hierarchical triple. The 40-yr
visual orbit with a semimajor axis slightly under 0\farcs40 is
reasonably well known \citep[e.g.,][]{vandenBos:1956,
  Soderhjelm:1999}, and numerous radial-velocity measurements of the
two visible stars have been collected mostly by \cite{Mermilliod:2009}
and \cite{Griffin:2012}, which now cover both the inner and outer
orbits well. However, so far it has not been possible to determine the
masses for all stars without making assumptions because the inner
binary is only single-lined.

Here we report our own, extensive spectroscopic monitoring of the
system for over three decades.  We have detected the weak lines of the
third star in most of our spectra and measured its velocities for the
first time, enabling the masses to be determined for all three stars.
This adds significantly to the small sample of Hyades binaries
available to establish the empirical MLR in the cluster. Furthermore,
we show that a global orbital solution combining our velocities and
those of others with the visual observations and other astrometric
constraints makes it possible to achieve much improved precision in the
masses compared to most of the other systems.

We present our spectroscopic observations and the velocity
measurements for the three stars in Section~\ref{sec:spectroscopy}.
The visual observations of the wide pair spanning more than a century
are described in Section~\ref{sec:astrometry}, together with other
measurements and astrometric constraints from the \hip\ and
\gaia\ missions that measure the acceleration in the plane of the
sky. Our orbital solution combining all measurements is reported in
Section~\ref{sec:analysis}. The updated empirical MLR of the Hyades is
discussed in Section~\ref{sec:MLR}, with a comparison to current
models of stellar evolution. Our closing remarks are in
Section~\ref{sec:remarks}.

%%%%%%%%%%%%%%%%%%%%%%%%%%%%%%%%%%%%%%%%%%%%%%%%%%%%%%%%%%%%%%%%%%%%%%%%
\section{Radial velocity measurements}
\label{sec:spectroscopy}
%%%%%%%%%%%%%%%%%%%%%%%%%%%%%%%%%%%%%%%%%%%%%%%%%%%%%%%%%%%%%%%%%%%%%%%%

The history of the radial-velocity observations of \vstar\ has been
documented by \cite{Griffin:2012}, including the initial confusion as
to which star of the visual pair was the 21-day spectroscopic
binary. That issue was finally resolved by \cite{Smekhov:1995}, who
showed that it is the visual secondary, and published the first radial
velocities for this star along with the elements of its spectroscopic
orbit.

\cite{Griffin:2012} compiled a list of velocity measurements resulting
from his own long-term monitoring of the system with several different
instruments, as well as those from a similar program by
\cite{Mermilliod:2009}. Together these 118 measurements span the
interval from January of 1972 to January of 2010. An effort was made
by Griffin to place them all on the same velocity system, so that
effectively they may be considered as a single data set. The
measurements are for the primary of the 40-yr visual pair (``star A'')
and the primary of the 21-day inner binary (``star Ba''). The
secondary of the 21-day binary will be referred to in the following as
``star Bb''. Where a brief designation is needed for the visual
secondary, it will be called ``star B''. Our orbital analysis of
Section~\ref{sec:analysis} below will make use of the Griffin
velocities as well as those of \cite{Smekhov:1995}, which in addition
to the 29 velocities for star Ba include 21 of star A. Several of the
early measurements in the list by \cite{Griffin:2012} were considered
by him to be unreliable, and we will reject them as well. His final
data set has 111 velocities for star A and 116 for star Ba.

Our own spectroscopic monitoring of \vstar\ at the Center for
Astrophysics (CfA), carried out as part of a large survey of the
Hyades cluster, began in November of 1980 and continued until
September of 2013. Observations were gathered with four different
instruments.  Spectra through November of 2007 were made with the CfA
Digital Speedometers \citep[DS;][]{Latham:1992} on the 1.5m
Tillinghast reflector at the Fred L.\ Whipple Observatory (Mount
Hopkins, AZ), the 1.5m Wyeth reflector at the Oak Ridge Observatory
(in the town of Harvard, MA), now closed, and the 4.5m-equivalent
Multiple Mirror Telescope (also on Mount Hopkins) before its
conversion to a monolithic 6.5m mirror. These echelle instruments with
a resolving power of $R \approx 35,000$ were equipped with intensified
photon-counting Reticon detectors and recorded a single order
45~\AA\ wide centered at 5187~\AA, featuring the \ion{Mg}{1}~b
triplet. Signal-to-noise ratios for the 169 usable DS spectra range
from about 20 to 83 per resolution element of 8.5~\kms, although at
the higher count levels flatfielding errors, rather than counts, are
likely to dominate the uncertainty. An additional 28 spectra of higher
quality were obtained using the Tillinghast Reflector Echelle
Spectrograph \citep[TRES;][]{Szentgyorgyi:2007, Furesz:2008}, a
bench-mounted fiber-fed echelle instrument attached to the 1.5m
Tillinghast reflector, with a resolving power of $R \approx
44,000$. These spectra cover the wavelength range 3800--9100~\AA\ in
51 orders. For the order centered at about 5187~\AA\ the
signal-to-noise ratios range between 81 and 284 per resolution element
of 6.8~\kms. Nightly sky exposures at dusk and dawn were used at each
telescope to monitor the zero point of the DS instruments and place
their velocities on a uniform system, which we refer to as the native
CfA system.
%
%Although not relevant for
% Although not necessary for mass determinations...
%our purposes, a small correction of +0.14~\kms\ should be added to the
%velocities on the native system to place them on the reference frame
%of the solar system minor planets \citep{Stefanik:1999, Latham:2002}.
%
For TRES we relied on IAU standard stars to transfer the velocities to
the same system as the DS.

We measured radial velocities using TRICOR, a three-dimensional
cross-correlation technique introduced by \cite{Zucker:1995} that uses
three templates, one for each component. We selected the templates
from a large library of synthetic spectra based on model atmospheres
by R.\ L.\ Kurucz, computed for the resolution of the DS and TRES
instruments \citep[see][]{Nordstrom:1994, Latham:2002}. These spectra
span 300~\AA\ centered on the \ion{Mg}{1}~b triplet, which therefore
includes the entire 45~\AA\ order of the DS spectra. For the velocity
determinations from TRES, we used only the 100~\AA\ order centered
around 5187~\AA, as experience shows that it contains most of the
velocity information.

The two main template parameters affecting the velocities are the
effective temperature ($T_{\rm eff}$) and rotational broadening ($v
\sin i$ when seen in projection).  For the surface gravity parameter
($\log g$) we adopted the value 4.5, which is appropriate for dwarf
stars such as those in \vstar, and for the metallicity parameter
([Fe/H]) we chose to use the solar value. This is within less than one
step in our metallicity grid of the true composition of the Hyades
\citep[${\rm [Fe/H]} = +0.18 \pm 0.03$;][]{Dutra-Ferreira:2016}, which
is sufficiently close for our purposes. After considerable
experimentation, we found the best results were obtained with template
temperatures of 6500~K, 6000~K, and 4500~K for stars A, Ba, and Bb,
respectively. The projected rotational velocity of star A has been
reported by \cite{Griffin:2012} to be $v_{\rm A} \sin i = 23.8 \pm
0.2~\kms$; we adopted 25~\kms, which is the nearest value in our
grid. Stars Ba and Bb were considered to be rotationally unbroadened,
which in the case of Ba agrees with the assessment of
\cite{Griffin:2012}.

\setlength{\tabcolsep}{6pt}  % tighten to make table fit in one column
\begin{deluxetable*}{lcccccccc}
\tablewidth{0pc}
\tablecaption{CfA Radial Velocities for \vstar\ \label{tab:rvs}}
\tablehead{
\colhead{HJD} &
\colhead{$RV_{\rm A}$} &
\colhead{$\sigma_{\rm A}$} &
\colhead{$RV_{\rm Ba}$} &
\colhead{$\sigma_{\rm Ba}$} &
\colhead{$RV_{\rm Bb}$} &
\colhead{$\sigma_{\rm Bb}$} &
\colhead{Inner} &
\colhead{Outer}
\\
\colhead{(2,400,000+)} &
\colhead{(\kms)} &
\colhead{(\kms)} &
\colhead{(\kms)} &
\colhead{(\kms)} &
\colhead{(\kms)} &
\colhead{(\kms)} &
\colhead{Orbital Phase} &
\colhead{Orbital Phase}
}
\startdata
 44560.8550  &  44.85  &  0.60  &  66.37  &  0.44  & $-$13.56\phs &   5.52  &  0.49240  &  0.13442 \\
 44956.8528  &  43.32  &  0.92  &  20.85  &  0.67  & 58.07  &   8.47  &  0.51897  &  0.76575 \\
 45245.8832  &  45.33  &  0.87  &  23.28  &  0.64  & 58.67  &   8.01  &  0.53837  &  0.36437 \\
 45336.8159  &  44.42  &  1.53  &  \phn9.94  &  1.12  & 66.53  &  14.02\phn  &  0.54447  &  0.64267 \\
 45337.8263  &  44.82  &  0.63  &  12.44  &  0.46  & 77.35  &   5.81  &  0.54454  &  0.69021
\enddata
\tablecomments{(This table is available in its entirety in machine-readable form.)}
\end{deluxetable*}
\setlength{\tabcolsep}{6pt}  % tighten to make table fit in one column

In a few of the weaker exposures from the DS, we were not able to
measure the velocities of star Bb reliably, and only those for stars A
and Ba are reported. The heliocentric radial velocities on the native
CfA system for the three stars are listed in Table~\ref{tab:rvs} along
with their uncertainties. There are 169 for stars A and Ba and 158 for
star Bb. We also measured the flux ratios between pairs of stars using
TRICOR, obtaining $\ell_{\rm A}/\ell_{\rm Ba} = 2.079 \pm 0.095$ and
$\ell_{\rm Bb}/\ell_{\rm Ba} = 0.049 \pm 0.007$ at a mean wavelength
of 5187~\AA.

\begin{figure}
\epsscale{1.15}
\plotone{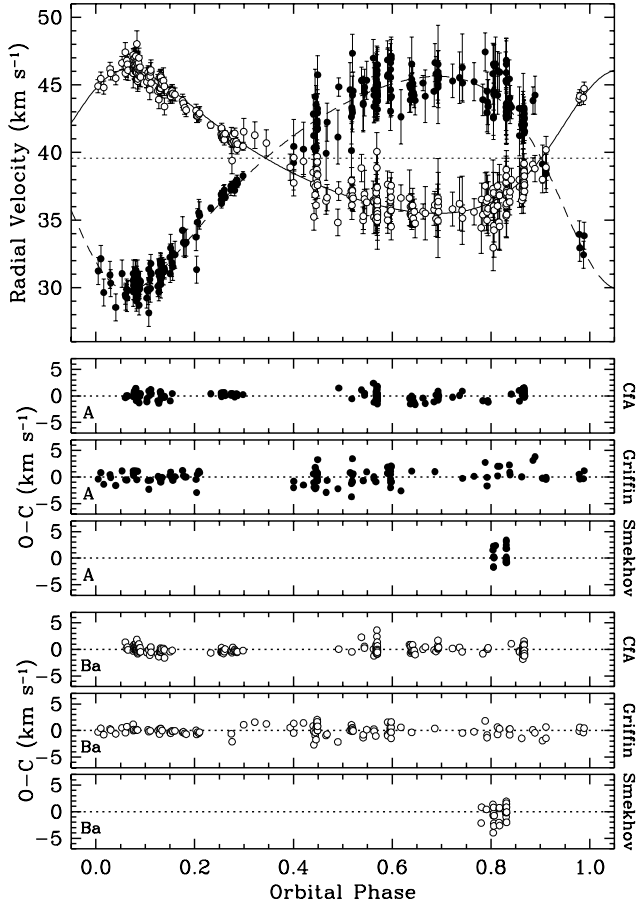}

\figcaption{\emph{Top panel:} Radial-velocity measurements of
  \vstar\ in the outer orbit, shown with our adopted model described
  in Section~\ref{sec:analysis}. The solid curve corresponds to the
  visual primary (star A), and the dashed one to the center of mass of
  the visual secondary (B). The points shown for the secondary are
  those of star Ba, corrected for motion in the inner orbit. We omit
  those of star Bb for clarity, as they show much larger scatter. The
  dotted line marks the center-of-mass velocity of the triple
  system. Measurements by \cite{Griffin:2012} and \cite{Smekhov:1995}
  have been adjusted for their respective zero-point offsets relative
  to CfA listed below in Table~\ref{tab:mcmc}. \emph{Bottom panels:}
  Residuals of the measurements for stars A and Bb from our adopted
  model, separately for the CfA, \cite{Griffin:2012}, and
  \cite{Smekhov:1995} observations.
\label{fig:RV_A}}

\end{figure}

\begin{figure}
\epsscale{1.15}
\plotone{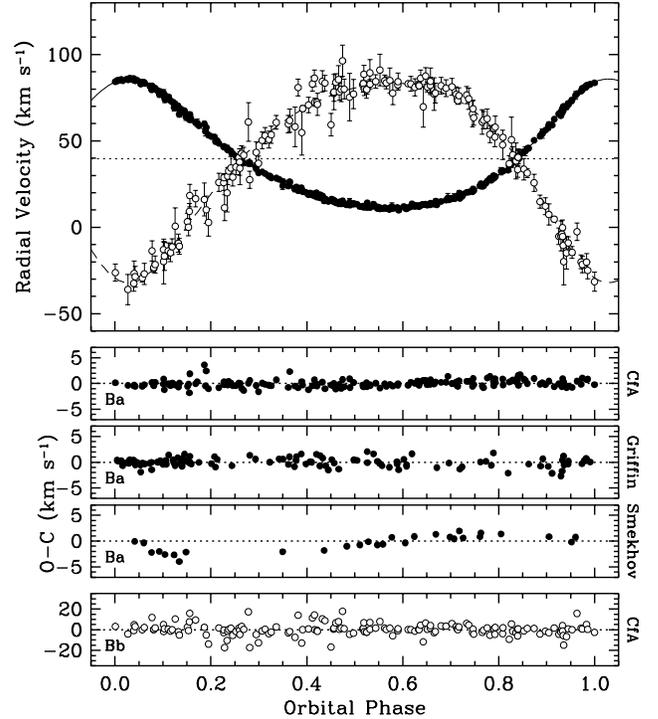}

\figcaption{\emph{Top panel:} Similar to Figure~\ref{fig:RV_A} for the
  inner orbit. The solid curve corresponds to star Ba and the dashed
  one to Bb. The observations shown for stars Ba and Bb have had the
  motion in the outer orbit removed. The dotted line marks the
  center-of-mass velocity of the triple system. \emph{Bottom panels:}
  Residuals of the measurements for the two stars, separately for the
  CfA, \cite{Griffin:2012}, and \cite{Smekhov:1995} observations.
\label{fig:RV_B}}

\end{figure}

Taken together, the CfA, \cite{Griffin:2012}, and \cite{Smekhov:1995}
data sets span a total of 36.5 years or about 90\% of the visual
binary period.  The velocity measurements from all data sets are shown
in Figure~\ref{fig:RV_A} and Figure~\ref{fig:RV_B} for the 40-yr outer
orbit and 21-day inner orbit, respectively, along with our adopted
model described later.

%%%%%%%%%%%%%%%%%%%%%%%%%%%%%%%%%%%%%%%%%%%%%%%%%%%%%%%%%%%%%%%%%%%%%%%%
\section{Astrometric observations}
\label{sec:astrometry}
%%%%%%%%%%%%%%%%%%%%%%%%%%%%%%%%%%%%%%%%%%%%%%%%%%%%%%%%%%%%%%%%%%%%%%%%

%%%%%%%%%%%%%%%%%%%%%%%%%%%%%%%%%%%%%%%%%%%%%%%%%%%%%%%%%%%%%%%%%%%%%%%%
\subsection{Visual measurements}
\label{sec:visual}
%%%%%%%%%%%%%%%%%%%%%%%%%%%%%%%%%%%%%%%%%%%%%%%%%%%%%%%%%%%%%%%%%%%%%%%%

Since its discovery in 1904 the relative positions of \vstar\ (angular
separations, $\rho$, and position angles, $\theta$) have been measured
by visual observers about 130 times, not counting another dozen
occasions in which the pair was not resolved. The most recent
observation was obtained at the end of 2016. Measurements until about
1970 were made with filar micrometers, and most observations since
were made using the speckle interferometry technique. A listing of all
measurements was kindly provided to us by Brian Mason (U.S.\ Naval
Observatory), extracted from the Washington Double Star Catalog (WDS),
with the dates of observation having been uniformly converted from the
traditional Besselian years to Julian years.

Most of these observations have no reported uncertainties. Their
quality varies greatly and depends on many factors including the
observing conditions, the telescope aperture used, and even the
experience and disposition of the observer. Assigning realistic errors
to any particular observation is non-trivial, and there is no unique
way of doing this. Here we have chosen to divide the observations into
groups by time period and by observational technique, and to assign
uniform errors within each group iterating during the analysis
described later to reach reduced $\chi^2$ values near unity within
each group. For the position angles we specified the uncertainties in
seconds of arc in the tangential direction, $\sigma_{\rm t}$, in order
to explicitly take into account the dependence of the angular
precision ($\sigma_{\theta}$, expressed in degrees) on the angular
separation: $\sigma_{\rm t} = \rho\thinspace \sigma_{\theta}$. For
observations prior to 1950 we adopted $\sigma_{\rho} = 0\farcs043$ and
$\sigma_{\rm t} = 0\farcs037$; micrometer measurements between 1950
and 1970 were assigned $\sigma_{\rho} = 0\farcs035$ and $\sigma_{\rm
  t} = 0\farcs024$; more recent micrometer measurements had
$\sigma_{\rho} = 0\farcs026$ and $\sigma_{\rm t} = 0\farcs030$; and
speckle observations received errors of $\sigma_{\rho} = 0\farcs004$
and $\sigma_{\rm t} = 0\farcs0034$. In four cases (observations on
1923.93, 1925.10, 1927.06, and 1928.17 we found that the quadrants of
the position angles needed to be reversed compared to the original
records from the WDS. In a few other cases, the observations (mostly in
angular separation) were rejected for being clearly different from
others near in time.

\begin{figure}
\epsscale{1.15}
\plotone{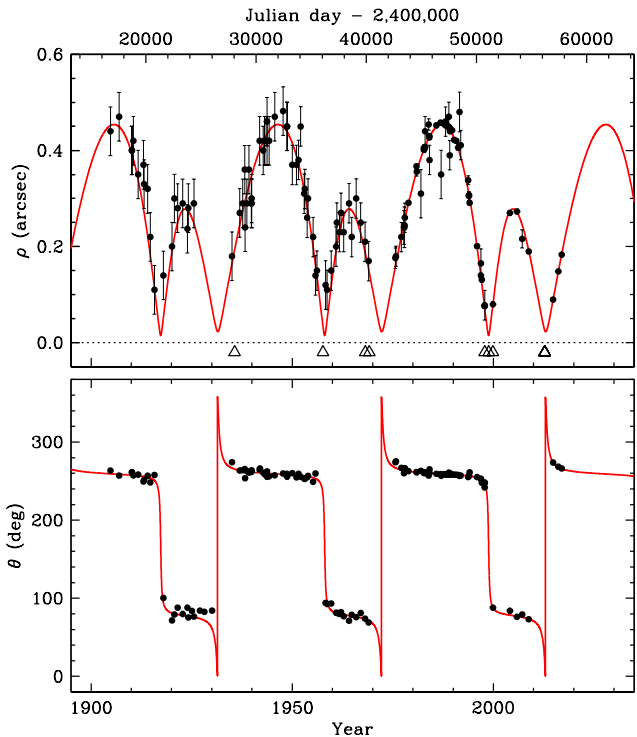}

\figcaption{Angular separation and position angle measurements
  ($\rho$, $\theta$) for \vstar\ as a function of time, with the
  adopted model described below in Section~\ref{sec:analysis}.
  Triangles at the bottom of the top panel indicate dates when the
  observers reported the binary was unresolved.
\label{fig:orbit}}

\end{figure}

In the end, we retained 126 measurements of the angular separation and
130 of the position angle of \vstar\ for the analysis below. These
observations cover slightly more than two and a half cycles of the
40-yr orbit and are shown in Figure~\ref{fig:orbit} together with our
adopted model described later. A listing of the measurements is given
in Table~\ref{tab:visual}.

\setlength{\tabcolsep}{6pt}  % tighten to make table fit in one column
\begin{deluxetable*}{lcccccccc}
\tablewidth{0pc}
\tablecaption{Visual Observations of \vstar \label{tab:visual}}
\tablehead{
\colhead{Year} &
\colhead{$\theta$ (\arcdeg)} &
\colhead{$\sigma_{\rm t}$ (\arcsec)} &
\colhead{$(O-C)_{\theta}$ (\arcdeg)} &
\colhead{$(O-C)_{\theta}/\sigma_{\theta}$} &
\colhead{$\rho$ (\arcsec)} &
\colhead{$\sigma_{\rho}$ (\arcsec)} &
\colhead{$(O-C)_{\rho}$ (\arcsec)} &
\colhead{$(O-C)_{\rho}/\sigma_{\rho}$}
%\\
%\colhead{} &
%\colhead{(deg)} &
%\colhead{(arcsec)} &
%\colhead{(deg)} &
%\colhead{} &
%\colhead{(arcsec)} &
%\colhead{(arcsec)} &
%\colhead{(arcsec)} &
%\colhead{}
}
\startdata
 1904.81   & 263.1  &  0.0238  & +4.26   & +1.41   &  0.44  &  0.0505  & $-$0.0118 & $-$0.23 \\
 1906.96   & 256.6  &  0.0238  & $-$1.59 & $-$0.52 &  0.47  &  0.0505  & +0.0205   & +0.41 \\
 1910.13   & 260.7  &  0.0238  & +3.59   & +1.04   &  0.40  &  0.0505  & +0.0065   & +0.13 \\
 1910.132  & 260.7  &  0.0238  & +3.59   & +1.04   &  0.40  &  0.0505  & +0.0066   & +0.13 \\
 1910.51   & 256.8  &  0.0238  & $-$0.15 & $-$0.04 &  0.42  &  0.0505  & +0.0379   & +0.75 
\enddata

\tablecomments{The position angles listed are the original ones;
  precession corrections were applied internally during the orbital
  analysis. The uncertainties listed in the table correspond to the
  $\sigma_{\rho}$ and $\sigma_{\rm t}$ values given in the text
  multiplied by the scaling factors $f_{\rho}$ and $f_{\theta}$
  reported in Section~\ref{sec:analysis}, and represent the final
  errors used in the solution described there. Also listed are the
  residuals in angular separation and position angle normalized to
  their uncertainties $\sigma_{\rho}$ and $\sigma_{\theta} =
  \sigma_{\rm t}/\rho$. (This table is available in its entirety in
  machine-readable form.)}

\end{deluxetable*}
\setlength{\tabcolsep}{6pt}  % default value

%%%%%%%%%%%%%%%%%%%%%%%%%%%%%%%%%%%%%%%%%%%%%%%%%%%%%%%%%%%%%%%%%%%%%%%%
\subsection{Lunar occultations}
\label{sec:lunar}
%%%%%%%%%%%%%%%%%%%%%%%%%%%%%%%%%%%%%%%%%%%%%%%%%%%%%%%%%%%%%%%%%%%%%%%%

\vstar\ underwent a series of occultation events by the Moon between
1978 and 1980, many of which were recorded by observers, sometimes in
several different filters. While these measurements are only
one-dimensional in nature and therefore do not provide as much
information as the visual measurements, they are usually much more
precise, they can have superior angular resolution, and typically also
yield a good measurement of the magnitude difference between the
components of a resolved binary. We, therefore, made use of these
measurements in our orbital analysis of Section~\ref{sec:analysis}. Of
the 11 recorded events included in the WDS listing we received, one by
\cite{Peterson:1981} was a near-grazing event and is best ignored, as
described therein and as recommended also by \citep{Evans:1984}.
Another by \cite{Richichi:1999} that was not in the WDS listing did
not resolve the pair; this may have been due to instrumental problems,
but we note that it was taken precisely at periastron, when the
predicted angular separation was only 15 milli-arc seconds. The
remaining 10 measurements are of good quality and were retained for
our analysis; they are listed in Table~\ref{tab:occ}.

\setlength{\tabcolsep}{4pt}  % tighten to make table fit in one column
\begin{deluxetable}{lccccc}
\tablewidth{0pc}
\tablecaption{Lunar Occultation Observations of \vstar \label{tab:occ}}
\tablehead{
\colhead{Year} &
\colhead{$\psi$} &
\colhead{$v$} &
\colhead{$\sigma_{v}$} &
\colhead{$(O-C)_{v}$} &
\colhead{$(O-C)_{v}/\sigma_{v}$}
\\
\colhead{} &
\colhead{(\arcdeg)} &
\colhead{(\arcsec)} &
\colhead{(\arcsec)} &
\colhead{(\arcsec)} &
\colhead{}
}
\startdata
 1978.7241  &  253.4 & 0.284\phn  &  0.0054  &  +0.0041   &  +0.76   \\
 1978.7241  &  254.5 & 0.258\phn  &  0.0134  &  $-$0.0227 &  $-$1.69 \\
 1978.7241  &  254.5 & 0.267\phn  &  0.0134  &  $-$0.0137 &  $-$1.02 \\
 1978.7242  &  241.2 & 0.2731     &  0.0089  &  +0.0092   &  +1.04   \\
 1979.1728  &  279.4 & 0.2794     &  0.0121  &  $-$0.0071 &  $-$0.59 \\
 1979.1728  &  286.3 & 0.2725     &  0.0030  &  $-$0.0014 &  $-$0.47 \\
 1979.1728  &  287.1 & 0.281\phn  &  0.0081  &  +0.0088   &  +1.10   \\
 1980.0702  &  259.2 & 0.332\phn  &  0.0188  &  +0.0025   &  +0.13   \\
 1980.0704  &  262.5 & 0.3400     &  0.0107  &  +0.0102   &  +0.95   \\
 1980.0705  &  258.9 & 0.3306     &  0.0016  &  +0.0012   &  +0.72  
\enddata

\tablecomments{The vector angles $\psi$ listed are the original ones;
  precession corrections were applied internally during the orbital
  analysis. The uncertainties in the vector separations $v$ are the
  original values multiplied by the scaling factor $f_{\rm occ}$
  reported in Section~\ref{sec:analysis}, and represent the final
  errors used in the solution described there. Residuals for the
  vector separations are listed in arc seconds and also normalized to
  their uncertainties.}
\end{deluxetable}

%%%%%%%%%%%%%%%%%%%%%%%%%%%%%%%%%%%%%%%%%%%%%%%%%%%%%%%%%%%%%%%%%%%%%%%%
\subsection{\hip\ and \gaia}
\label{sec:hip}
%%%%%%%%%%%%%%%%%%%%%%%%%%%%%%%%%%%%%%%%%%%%%%%%%%%%%%%%%%%%%%%%%%%%%%%%

\vstar\ has entries in both the \hip\ Catalogue \citep{ESA:1997} and
in the second data release of the \gaia\ Catalogue
\citep[\gaia/DR2;][]{Gaia:2018}. The respective identifiers are
HIP\thinspace 20916 and \gaia/DR2 3312892757136810880. The proper
motions (p.m.)\ measured at each epoch (circa 1991.25 and 2015.5)
differ significantly, which is a reflection of the acceleration in the
plane of the sky due to the changing positions in the 40-yr orbit. The
p.m.\ difference therefore contains information on the orbit, and can
be used to constrain it, as we do below.  \cite{Brandt:2018} published
a cross-calibrated catalog of \hip\ and \gaia\ astrometry that
facilitates this, and includes for each star a third p.m.\ measurement
derived from the positional difference between the two original
catalogs divided by the $\sim$24~yr time baseline. As demonstrated by
\cite{Brandt:2018} this last p.m.\ measure is practically independent
of the other two and is particularly useful as it is considerably more
precise.

We used all three p.m.\ determinations in this work to supplement the
visual and lunar occultation observations and improve the orbital
elements. At the time of the \hip\ mission the angular separation
between \vstar~A and B was about 0\farcs4, and the instrument was able
to resolve the pair; the corresponding $\rho$ and $\theta$
measurements are included in the listing from the WDS. The published
position and p.m.\ correspond to the primary, as indicated in the
Catalogue. On the other hand, in 2015.5 the separation was only about
0\farcs12, which is essentially at the resolution limit of
\gaia\ \citep{Gaia:2016}. Values published in \gaia/DR2, therefore,
correspond to the center of light rather than the primary. This
important detail will be accounted for below.

\setlength{\tabcolsep}{6pt}  % tighten to make table fit in one column
\begin{deluxetable*}{lcccccc}
\tablewidth{0pc}
\tablecaption{Proper Motion and Parallax Information for \vstar\ from
  \gaia/DR2 and \hip \label{tab:pm}}
\tablehead{
\colhead{Source} &
\colhead{$\mu_{\alpha}^*$ (mas yr$^{-1}$)} &
\colhead{$\mu_{\delta}$ (mas yr$^{-1}$)} &
\colhead{Corr} &
\colhead{Average Epoch} &
\colhead{$\pi$ (mas)}
}
\startdata
HIP        & \phn$+89.35 \pm 1.53$    & $-24.08 \pm 0.95$   & $-0.098$ & 1991.26 / 1990.95 & $20.62 \pm 1.27$ \\
\gaia--HIP & \phn$+96.783 \pm 0.071$  & $-26.944 \pm 0.035$ & $+0.325$ & \nodata & \nodata \\
\gaia      & $+117.11 \pm 0.62$   & $-22.74 \pm 0.39$   & $-0.206$ & 2015.58 / 2015.61 & $20.74 \pm 0.18$
\enddata

\tablecomments{Entries are taken from the acceleration catalog of
  \cite{Brandt:2018}. $\mu_{\alpha}^*$ represents the p.m.\ in
  R.A.\ multiplied by the cosine of the Declination, and ``Corr'' is
  the correlation coefficient between the proper motions in R.A.\ and
  Dec. The ``Average Epoch'' is given separately for the
  p.m.\ measurements in R.A.\ and Dec. The \hip\ parallax is a
  combination of results from the original mission and the
  re-reduction \citep{ESA:1997, vanLeeuwen:2007} (see text).}

\end{deluxetable*}
\setlength{\tabcolsep}{6pt}  % default value

Table~\ref{tab:pm} collects these measurements taken directly from the
catalog of \cite{Brandt:2018}, along with the uncertainties and
correlation coefficients as listed there, and the average time of
observation for each catalog in each coordinate. We include also the
parallax measurements from the two missions, which we used as well to
constrain our orbital fit as described below. The value for \hip\ is
the 60/40 combination of the results from the re-reduction performed
by \cite{vanLeeuwen:2007} and from the original \hip\ catalog
\citep{ESA:1997}, following \cite{Brandt:2018}, and includes an error
inflation of 0.2~mas in quadrature as prescribed in his Eq.(18).

%%%%%%%%%%%%%%%%%%%%%%%%%%%%%%%%%%%%%%%%%%%%%%%%%%%%%%%%%%%%%%%%%%%%%%%%
\section{Analysis and Results}
\label{sec:analysis}
%%%%%%%%%%%%%%%%%%%%%%%%%%%%%%%%%%%%%%%%%%%%%%%%%%%%%%%%%%%%%%%%%%%%%%%%

We combined all radial-velocity and astrometric measurements into a
global solution solving simultaneously for the elements of the inner
and outer orbits, which were assumed to be non-interacting. The visual
orbit (AB) has the following parameters: the period $P_{\rm AB}$, the
angular semimajor axis $a^{\prime\prime}_{\rm AB}$, eccentricity
parameters $\sqrt{e_{\rm AB}}\cos\omega_{\rm B}$ and $\sqrt{e_{\rm
    AB}}\sin\omega_{\rm B}$ (where $e_{\rm AB}$ is the eccentricity
and $\omega_{\rm B}$ the argument of periastron for star B, following
the visual binary convention), the cosine of the inclination angle
$\cos i_{\rm AB}$, the position angle of the ascending node
$\Omega_{\rm AB}$ referred to the equinox J2000, and a reference time
of periastron passage $T_{\rm AB}$. Precession corrections to J2000
were applied to all position angles from the visual and speckle
measurements, as well as to the vector angles $\psi$ from the lunar
occultations.

The spectroscopic elements of the outer orbit include the
center-of-mass velocity $\gamma$ of the triple, and the velocity
semiamplitudes of each visual component, $K_{\rm A}$ and $K_{\rm B}$.
The inner spectroscopic orbit is described by its period $P_{\rm B}$,
the corresponding eccentricity parameters $\sqrt{e_{\rm
    B}}\cos\omega_{\rm Ba}$ and $\sqrt{e_{\rm B}}\sin\omega_{\rm Ba}$
(with $\omega_{\rm Ba}$ being the argument of periastron for star Ba,
following the usual spectroscopic convention), the velocity
semiamplitudes $K_{\rm Ba}$ and $K_{\rm Bb}$, and a reference time of
periastron passage $T_{\rm B}$.  Two extra parameters were included to
account for possible zero-point offsets between the
\cite{Griffin:2012} and \cite{Smekhov:1995} velocities and our own:
$\Delta RV_{\rm G}$ and $\Delta RV_{\rm S}$.  These shifts are to be
added to those data sets in order to bring them onto the native CfA
system. Though barely significant, light travel time corrections were
computed at each step during the iterations and applied to the times
of observation for the inner orbit. The maximum effect is about 0.047
days, corresponding to 0.0022 in phase.

The formalism for incorporating the \hip, \gaia, and
\hip--\gaia\ proper motions ($\mu^*_{\alpha}$ and
$\mu_{\delta}$)\footnote{The $\mu^*_{\alpha}$ notation represents the
  p.m.\ in Right Ascension multiplied by $\cos\delta$.} was described
in detail by \cite{Torres:2019} \citep[see also][]{Brandt:2018,
  Dupuy:2019, Brandt:2019}, and we refer the reader to that work for
details. Briefly, and using \hip\ as an example, the p.m.\ in
R.A.\ measured by that mission for the primary star
($\mu^*_{\alpha,\rm H,A}$) is simply that of the barycenter of the
triple system ($\mu^*_{\alpha,0}$) plus a perturbation from the
orbital motion of the primary around the barycenter:
$\mu^*_{\alpha,\rm H,A} = \mu^*_{\alpha,0} + \Delta\mu^*_{\alpha,\rm
  H,A}$. A similar expression can be written for the Declination
component. The first term on the right is an adjustable parameter in
our analysis (along with $\mu_{\delta,0}$), and the second represents
the change with time of the orbital position of the primary relative
to the barycenter, at the average epoch $t_{\alpha,\rm H}$ of the
\hip\ measurement, as given in Table~\ref{tab:pm}. Although both
\hip\ and \gaia\ observed \vstar\ over finite intervals of time of 2.5
and 1.7 yr, respectively, these are relatively short compared to the
orbital period. We, therefore, assumed $\Delta\mu^*_{\alpha,\rm H,A}$
(and an analogous perturbation term for \gaia) to be an instantaneous
quantity and computed it as the time derivative of the orbital
position.

Similarly, the \hip--\gaia\ proper motion in R.A., $\mu^*_{\alpha,\rm
  HG}$, is that of the center of mass of the triple with an added term
that incorporates the change in position between the two epochs:
$$\mu^*_{\alpha,\rm HG} = \mu^*_{\alpha,0} +
\frac{\Delta\alpha^*_{\rm G}[t_{\alpha,\rm G}] - \Delta\alpha^*_{\rm
    H}[t_{\alpha,\rm H}]}{t_{\alpha,\rm G} - t_{\alpha,\rm H}}.$$
Here the $\Delta\alpha^*$ quantities represent the position measured
by the two missions relative to the barycenter at the mean epoch of
each catalog and can be calculated from the orbital elements listed
above. A similar equation holds for the Declination component
$\mu_{\delta,{\rm HG}}$.

A complicating factor in the above is that \hip\ measured the position
and p.m.\ of the primary star, whereas \gaia\ measured the
photocenter, as mentioned previously.  The position and motion of star
A can be computed easily enough from the orbital elements listed
above, but doing the same for the center of light in the case of
\gaia\ requires us to know the semimajor axis of the AB photocenter,
$a^{\prime\prime}_{\rm ph, AB}$. The relation between semimajor axis
of the photocenter and that of the orbit of star B relative to A
($a^{\prime\prime}_{\rm AB}$) is $a^{\prime\prime}_{\rm ph,AB} =
a^{\prime\prime}_{\rm AB} (f_{\rm AB} - \beta_{\rm G})$, in which
$f_{\rm AB}$ is the fractional mass in the outer orbit and $\beta_{\rm
  G}$ is the fractional light in the \gaia\ bandpass. The mass
fraction $f_{\rm AB} \equiv M_{\rm B}/(M_{\rm A} + M_{\rm B})$ can be
recast in terms of the orbital elements as $f_{\rm AB} = K_{\rm
  A}/(K_{\rm A} + K_{\rm B})$. The fractional light $\beta_{\rm G}$ is
unknown a priori but is constrained by the astrometry, so we included
it as an additional adjustable parameter. This enables us to properly
model the proper motions involving \gaia.

%The light fraction can be expressed as $\beta_{\rm G} =
%1/(1+10^{+0.4 \Delta G})$, where $\Delta G$ is the magnitude
%difference between stars B and A at the \gaia\ wavelengths. In
%summary, to be able to use the \gaia\ and \hip--\gaia\ proper motions
%for this application we must first determine $\Delta G$, even though
%\gaia\ did not resolve the visual pair. The next section describes how
%we did this, leading to the estimate $\beta_{\rm G} = 0.355 \pm
%0.004$~mag that we use here.

We carried out our analysis within a Markov Chain Monte Carlo (MCMC)
framework using the {\tt emcee\/}\footnote{\url
  https://emcee.readthedocs.io/en/v2.2.1/} code of
\cite{Foreman-Mackey:2013}, which is a Python implementation of the
affine-invariant MCMC ensemble sampler proposed by
\cite{Goodman:2010}. We used 100 walkers with 10,000 links each, after
discarding the burn-in. Priors for most variables were uniform, with
those for $P_{\rm AB}$ and $a^{\prime\prime}_{\rm AB}$ being
log-uniform. Convergence was checked by visual inspection of the
chains, requiring also a Gelman-Rubin statistic \citep{Gelman:1992}
less than 1.05 for all adjustable parameters.

The use of different kinds of observations requires careful relative
weighting for a balanced solution. We handled this by incorporating
multiplicative scaling factors for the uncertainties, which we
adjusted simultaneously and self-consistently with the other orbital
parameters \citep[see][]{Gregory:2005}. For the CfA velocities, the
initial uncertainties for the three stars are those listed in
Table~\ref{tab:rvs}. Our analysis used a different error inflation
factor for each star, $f_{\rm CfA,A}$, $f_{\rm CfA,Ba}$, and $f_{\rm
  CfA,Bb}$. The uncertainties for the \cite{Smekhov:1995} velocities
were adopted as published, and the corresponding error inflation
factors we introduced are $f_{\rm S,A}$ and $f_{\rm S,Ba}$. For the
\cite{Griffin:2012} velocities we took into account the relative
weighting of the observations recommended by the author for each star
and each instrument and used the reported error of unit weight to
calculate the formal uncertainties. The corresponding error scaling
factors we then added are $f_{\rm G,A}$ and $f_{\rm G,Ba}$. We
proceeded similarly with the angular separations and position angles
from the visual measurements, and with the lunar occultations, adding
three more error scaling factors $f_{\rho}$, $f_{\theta}$, and $f_{\rm
  occ}$. Here $f_{\theta}$ represents the error inflation factor for
the P.A.\ uncertainties in the tangential direction, $\sigma_{\rm t}$.
The adopted priors for all of these factors were log-uniform. The
uncertainties for the p.m.\ in Table~\ref{tab:pm} were taken at face
value, and correlations between the R.A.\ and Dec.\ components were
accounted for as described by \cite{Torres:2019} in computing the
likelihood function.

%As a final adjustable parameter in our analysis we added $\beta_{\rm
%  G}$, which is constrained to some extent by the simultaneous use of
%the different kinds of observations. The constraint is weak, however,
%so we chose to use our estimate of this quantity mentioned earlier as
%a Gaussian prior. 

We note that the combination of the visual and spectroscopic
observations of \vstar\ yields the orbital parallax of the system. As
\hip\ and \gaia\ provided independent trigonometric parallaxes, we
used those determinations as measurements with their corresponding
uncertainties reported in Table~\ref{tab:pm} to further constrain the
solution.

\setlength{\tabcolsep}{4pt}
\begin{deluxetable}{lcc}
\tablewidth{0pc}
\tablecaption{Results of our MCMC Analysis for \vstar \label{tab:mcmc}}
\tablehead{ \colhead{~~~~~~~~~~~Parameter~~~~~~~~~~~} & \colhead{Value} & \colhead{Prior} }
\startdata
 $P_{\rm AB}$ (year)\dotfill                   & $40.752^{+0.031}_{-0.034}$  & [1, 5]* \\ [1ex]
 $a^{\prime\prime}_{\rm AB}$ (arcsec)\dotfill  & $0.3749^{+0.0010}_{-0.0010}$  & [$-$3, 2]* \\ [1ex]
 $\sqrt{e_{\rm AB}}\cos\omega_{\rm B}$\dotfill & $+0.4154^{+0.0027}_{-0.0027}$ & [$-$1, 1] \\ [1ex]
 $\sqrt{e_{\rm AB}}\sin\omega_{\rm B}$\dotfill & $-0.3811^{+0.0053}_{-0.0053}$ & [$-$1, 1] \\ [1ex]
 $\cos i_{\rm AB}$\dotfill                     & $-0.0534^{+0.0012}_{-0.0014}$     & [$-$1, 1] \\ [1ex]
 $\Omega_{\rm AB}$ (degree)\dotfill            & $78.395^{+0.059}_{-0.060}$         & [$-$180, +180] \\ [1ex]
 $T_{\rm AB}$ (year)\dotfill                   & $1960.845^{+0.070}_{-0.070}$ & [1900, 2000] \\ [1ex]
 $\gamma$ (\kms)\dotfill                       & $+39.582^{+0.038}_{-0.038}$ & [+20, +60  ] \\ [1ex]
 $K_{\rm A}$ (\kms)\dotfill                    & $7.927^{+0.064}_{-0.063}$ & [0 , 80] \\ [1ex]
 $K_{\rm B}$ (\kms)\dotfill                    & $5.343^{+0.057}_{-0.052}$ & [0 , 80] \\ [1ex]
 $P_{\rm B}$ (day)\dotfill                     & $21.254396^{+0.000024}_{-0.000025}$ & [20, 23] \\ [1ex]
 $\sqrt{e_{\rm B}}\cos\omega_{\rm Ba}$\dotfill & $+0.4794^{+0.0015}_{-0.0015}$ & [$-$1, 1] \\ [1ex]
 $\sqrt{e_{\rm B}}\sin\omega_{\rm Ba}$\dotfill & $-0.1530^{+0.0026}_{-0.0026}$ & [$-$1, 1] \\ [1ex]
 $K_{\rm Ba}$ (\kms)\dotfill                   & $37.245^{+0.054}_{-0.055}$ & [0, 80] \\ [1ex]
 $K_{\rm Bb}$ (\kms)\dotfill                   & $57.72^{+0.44}_{-0.41}$ & [0, 80] \\ [1ex]
 $T_{\rm B}$ (HJD$-2,400,000$)\dotfill         & $50828.045^{+0.016}_{-0.016}$ & [50800, 50850] \\ [1ex]
 $\Delta RV_{\rm G}$ (\kms)\dotfill            & $-0.849^{+0.072}_{-0.072}$ & [$-5$, +5] \\ [1ex]
 $\Delta RV_{\rm S}$ (\kms)\dotfill            & $+0.52^{+0.24}_{-0.24}$ & [$-5$, +5] \\ [1ex]
 $\mu^*_{\alpha,0}$ (mas yr$^{-1}$)\dotfill    & $+105.15^{+0.13}_{-0.13}$ & [+80, +140] \\ [1ex]
 $\mu_{\delta,0}$ (mas yr$^{-1}$)\dotfill      & $-24.861^{+0.040}_{-0.039}$ & [$-50$, 0] \\ [1ex]
 $\beta_{\rm G}$\dotfill                       & $0.336^{+0.014}_{-0.014}$ & [0.2, 0.5] \\ [1ex]
 $f_{\rm CfA,A}$\dotfill                       & $1.131^{+0.075}_{-0.064}$ & [$-5$, +3]* \\ [1ex]
 $f_{\rm CfA,Ba}$\dotfill                      & $1.581^{+0.111}_{-0.095}$ & [$-5$, +3]* \\ [1ex]
 $f_{\rm CfA,Bb}$\dotfill                      & $0.834^{+0.052}_{-0.044}$ & [$-5$, +3]* \\ [1ex]
 $f_{\rm G,A}$\dotfill                         & $1.073^{+0.077}_{-0.066}$ & [$-5$, +3]* \\ [1ex]
 $f_{\rm G,Ba}$\dotfill                        & $1.353^{+0.105}_{-0.090}$ & [$-5$, +3]* \\ [1ex]
 $f_{\rm S,A}$\dotfill                         & $2.84^{+0.56}_{-0.40}$ & [$-5$, +3]* \\ [1ex]
 $f_{\rm S,Ba}$\dotfill                        & $3.63^{+0.62}_{-0.45}$ & [$-5$, +3]* \\ [1ex]
 $f_{\rho}$\dotfill                            & $1.175^{+0.084}_{-0.071}$ & [$-5$, +3]* \\ [1ex]
 $f_{\theta}$\dotfill                          & $0.642^{+0.043}_{-0.037}$ & [$-5$, +3]* \\ [1ex]
 $f_{\rm occ}$\dotfill                         & $2.64^{+0.85}_{-0.49}$ & [$-5$, +3]*
\enddata
\tablecomments{The values listed correspond to the mode of the
  posterior distributions, and the uncertainties represent the 68.3\%
  credible intervals. Priors marked with an asterisk are log-uniform
  over the specified ranges; all others are uniform.}
\end{deluxetable}
\setlength{\tabcolsep}{6pt}

The results of the analysis are presented in Table~\ref{tab:mcmc},
with an indication of the priors we used. The model has a total of 31
adjustable parameters. We report the mode of the corresponding
posterior distributions for each parameter, along with the 68.3\%
confidence intervals. Table~\ref{tab:deriv} lists several derived
quantities computed directly from the chains of the corresponding
adjustable parameters involved. From the fitted value of $\beta_{\rm
  G}$, we infer a magnitude difference between the two visual
components in the \gaia\ bandpass of $\Delta G = 0.739 \pm 0.070$~mag.

\setlength{\tabcolsep}{4pt}
\begin{deluxetable}{lc}
\tablewidth{0pc}
\tablecaption{Derived Properties from our MCMC Analysis of \vstar \label{tab:deriv}}
\tablehead{ \colhead{~~~~~~~~~~~Quantity~~~~~~~~~~~} & \colhead{Value}}
\startdata
 $e_{\rm AB}$\dotfill                          & $0.3179^{+0.0024}_{-0.0024}$   \\ [1ex]
 $\omega_{\rm B}$ (degree)\dotfill             & $317.46^{+0.57}_{-0.57}$  \\ [1ex]
 $i_{\rm AB}$ (degree)\dotfill                 & $93.060^{+0.078}_{-0.071}$  \\ [1ex]
 $a_{\rm AB}$ (au)\dotfill                     & $17.689^{+0.095}_{-0.094}$   \\ [1ex]
 $a^{\prime\prime}_{\rm B}$ (mas)\dotfill      & $4.108^{+0.015}_{-0.015}$   \\ [1ex]
 $e_{\rm B}$\dotfill                           & $0.2532^{+0.0014}_{-0.0014}$   \\ [1ex]
 $\omega_{\rm Ba}$ (degree)\dotfill            & $342.30^{+0.31}_{-0.31}$  \\ [1ex]
 $i_{\rm B}$\tablenotemark{a} (degree)\dotfill & $75.1^{+1.6}_{-1.5}$  \\ [1ex]
 $a_{\rm B}$ (au)\dotfill                      & $0.1889^{+0.0011}_{-0.0011}$   \\ [1ex]
 $\pi_{\rm orb}$ (mas)\dotfill                 & $21.75^{+0.11}_{-0.11}$  \\ [1ex]
 Distance (pc)\dotfill                         & $45.98^{+0.23}_{-0.24}$    \\ [1ex]
 $M_{\rm A}$ ($M_{\sun}$)\dotfill              & $1.341^{+0.026}_{-0.024}$  \\ [1ex]
 $M_{\rm Ba}$ ($M_{\sun}$)\dotfill             & $1.210^{+0.021}_{-0.021}$  \\ [1ex]
 $M_{\rm Bb}$ ($M_{\sun}$)\dotfill             & $0.781^{+0.014}_{-0.014}$  \\ [1ex]
 $M_{\rm B}$ ($M_{\sun}$)\dotfill              & $1.991^{+0.034}_{-0.034}$  \\ [1ex]
 $q_{\rm AB} \equiv M_{\rm B}/M_{\rm A}$\dotfill & $1.483^{+0.022}_{-0.022}$  \\ [1ex]
 $q_{\rm B} \equiv M_{\rm Bb}/M_{\rm Ba}$\dotfill & $0.6452^{+0.0048}_{-0.0048}$ \\ [1ex]
 $\Delta G$ (mag)\dotfill                      & $0.739^{+0.070}_{-0.070}$  \\ [-1.5ex]
\tablecomments{The values listed correspond to the mode of the
  posterior distributions and the 68.3\% credible intervals.}
\tablenotetext{a}{Because the position angle of the ascending node is
  not known for the B orbit, the true inclination may also be on the
  opposite side of 90\arcdeg, i.e., 104\fdg9.}
\end{deluxetable}
\setlength{\tabcolsep}{6pt}

The spectroscopic and visual observations together with the adopted
model may be seen in Figures~\ref{fig:RV_A}, \ref{fig:RV_B}, and
\ref{fig:orbit} presented earlier. In Figure~\ref{fig:RV_B} we note
that the residuals of the \cite{Smekhov:1995} velocities for star Ba
show a clear sinusoidal pattern as a function of phase in the inner
orbit, indicative of some systematic error not present in the other
data sets. We examined this further by performing separate,
single-lined orbital solutions for the inner orbit using the Ba
velocities from CfA, \cite{Griffin:2012}, and \cite{Smekhov:1995}, all
corrected in the same way for motion in the outer orbit using the
parameters from the solution above. We found the three velocity
semi-amplitudes $K_{\rm Ba}$ to be consistent with each other within
uncertainties, so we chose to keep the \cite{Smekhov:1995} velocities
for our global analysis. The residuals of the astrometric observations
(visual and lunar occultation measurements) are given in
Table~\ref{tab:visual} and Table~\ref{tab:occ}, in natural units and
also normalized to the measurement uncertainties.

\begin{figure}
\epsscale{1.15}
\plotone{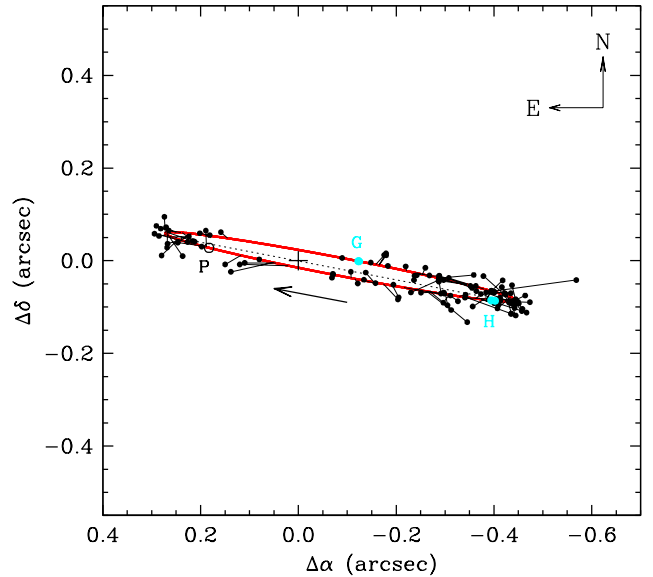}

\figcaption{Visual observations and adopted model for the 40-yr outer
  orbit in the plane of the sky. The plus sign marks the position of
  the primary. The direction of motion is clockwise, as indicated by
  the arrow. The dotted line is the line of nodes, and the open circle
  on the left labeled ``P'' indicates the location of periastron. Thin
  lines connect the observations with the predicted position along the
  orbit. For reference, we show also the position of the secondary at
  the time of the \hip\ (``H'') and \gaia\ (``G'') measurements.
\label{fig:orbitsky}}

\end{figure}

A representation of the outer orbit in polar coordinates is presented
in Figure~\ref{fig:orbitsky}, showing it to be nearly edge-on ($i_{\rm
  AB} \approx 92\fdg8$). The inner orbit is also quite highly
inclined. Table~\ref{tab:deriv} lists it as $i_{\rm B} \approx
75\arcdeg$, but it could also be the symmetrical value relative to
90\arcdeg\ (105\arcdeg), given that we are missing information on
$\Omega_{\rm B}$. Either way the two orbits cannot be exactly
coplanar.  The angular semimajor axis of the orbit of B is computed to
be $a^{\prime\prime}_{\rm B} \approx 4.1$~mas.  Careful re-examination
of the epoch astrometry from the \hip\ mission (transit observations)
with the improved knowledge we now have of the orbit of B did not
reveal the signature of that star, indicating those observations are
not quite precise enough to detect the wobble of stars Ba and Bb.
However, with a typical separation of the order of 4.1~mas spatial
resolution of the B pair should be within reach of modern
interferometers provided they have sufficient sensitivity. We estimate
the apparent $K$-band magnitude of the B pair to be about 6.0, and
that of the individual components as 6.2 (Ba) and 7.7 (Bb). The
angular separation of the Ba+Bb pair can now be predicted accurately
for any given time from the above elements; however, we cannot do the
same for the position angle because the orientation of the line of
nodes is unknown.

A comparison of our results with previously published orbital
solutions for the visual pair and the inner binary by
\cite{Smekhov:1995}, \cite{Soderhjelm:1999}, and \cite{Griffin:2012}
is presented in Table~\ref{tab:comparison}.

\setlength{\tabcolsep}{6pt}
\begin{deluxetable*}{lccccc}
\tablewidth{0pc}
\tablecaption{Comparison of Orbital Solutions for \vstar \label{tab:comparison}}
\tablehead{
\colhead{~~~~~~~~~~~~Parameter~~~~~~~~~~~~} &
\colhead{\cite{Smekhov:1995}} &
\colhead{\cite{Soderhjelm:1999}} &
\colhead{\cite{Griffin:2012}} &
\colhead{This work}
}
\startdata
\multicolumn{6}{c}{Outer orbit} \\ [0.5ex]
\hline \\ [-1ex]
 $P_{\rm AB}$ (year)\dotfill                   &  $40.38 \pm 0.01$\phn                  &  40.7    &  $40.24 \pm 0.58$\phn                     & $40.752^{+0.031}_{-0.034}$  \\ [1ex]          
 $a^{\prime\prime}_{\rm AB}$ (arcsec)\dotfill  &  $0.411 \pm 0.004$                     &  0.39    &  \nodata                                  & $0.3749^{+0.0010}_{-0.0010}$  \\ [1ex]        
 $e_{\rm AB}$\dotfill                          &  $0.39 \pm 0.01$                       &  0.33    &  $0.312 \pm 0.014$                        & $0.3179^{+0.0024}_{-0.0024}$   \\ [1ex]       
 $i_{\rm AB}$ (degree)\dotfill                 &  $92.5 \pm 0.4$\phn                    &  92      &  \nodata                                  & $93.060^{+0.078}_{-0.071}$  \\ [1ex]          
 $\omega_{\rm B}$ (degree)\dotfill             &  $296.88 \pm 0.09$\phn\phn             &  311     &  $315.0 \pm 3.0$\phn\phn                  & $317.46^{+0.57}_{-0.57}$  \\ [1ex]            
 $\Omega_{\rm AB}$ (degree)\dotfill            &  $78.0 \pm 0.2$\phn                    &  78      &  \nodata                                  & $78.395^{+0.059}_{-0.060}$  \\ [1ex]          
 $T_{\rm AB}$ (year)\dotfill                   &  \phm{*}$1958.12 \pm 0.02$*\phm{222}   &  1960    &  \phm{*}$1961.15 \pm 0.25$*\phm{222}      & $1960.845^{+0.070}_{-0.070}$  \\ [1ex]        
 $\gamma$ (\kms)\dotfill                       &  \nodata                               &  \nodata &  $+40.50 \pm 0.06$\phn\phs                & $+39.582^{+0.038}_{-0.038}$  \\ [1ex]         
 $K_{\rm A}$ (\kms)\dotfill                    &  \nodata                               &  \nodata &  $8.27 \pm 0.15$                          & $7.927^{+0.064}_{-0.063}$  \\ [1ex]           
 $K_{\rm B}$ (\kms)\dotfill                    &  \nodata                               &  \nodata &  $5.68 \pm 0.10$
                          & $5.343^{+0.057}_{-0.052}$  \\ [1ex]
\hline \\ [-1.5ex]
\multicolumn{6}{c}{Inner orbit} \\ [0.5ex]
\hline \\ [-1ex]
 $P_{\rm B}$ (day)\dotfill                     &  $21.253 \pm 0.001$\phn                &  \nodata &  $21.254259 \pm 0.000033$\phn             & $21.254396^{+0.000024}_{-0.000025}$  \\ [1ex] 
 $K_{\rm Ba}$ (\kms)\dotfill                   &  $37.04 \pm 0.13$\phn                  &  \nodata &  $37.07 \pm 0.10$\phn                     & $37.245^{+0.054}_{-0.055}$ \\ [1ex]           
 $e_{\rm B}$\dotfill                           &  $0.268 \pm 0.005$                     &  \nodata &  $0.2553 \pm 0.0019$                      & $0.2532^{+0.0014}_{-0.0014}$  \\ [1ex]        
 $\omega_{\rm Ba}$ (degree)\dotfill            &  $348.7 \pm 0.9$\phn\phn               &  \nodata &  $341.5 \pm 0.6$\phn\phn                  & $342.30^{+0.31}_{-0.31}$  \\ [1ex]            
 $T_{\rm B}$ (HJD$-2,400,000$)\dotfill         &  \phm{*}$50828.16 \pm 0.09$*\phm{2222} &  \nodata &  \phm{*}$50828.001 \pm 0.030$*\phm{2222}  & $50828.045^{+0.016}_{-0.016}$  \\ [-1.5ex]
\tablecomments{No uncertainties were reported for the orbital elements
  of \cite{Soderhjelm:1999}. Times of periastron passage marked with
  an asterisk are projected forward or backward by an integer number
  of cycles from the original times published to facilitate comparison
  with this work.}
\end{deluxetable*}
\setlength{\tabcolsep}{6pt}

%%%%%%%%%%%%%%%%%%%%%%%%%%%%%%%%%%%%%%%%%%%%%%%%%%%%%%%%%%%%%%%%%%%%%%%%
\section{Empirical mass-luminosity relation}
\label{sec:MLR}
%%%%%%%%%%%%%%%%%%%%%%%%%%%%%%%%%%%%%%%%%%%%%%%%%%%%%%%%%%%%%%%%%%%%%%%%

In order to make use of the mass determinations of \vstar\ from the
previous section to place them on the empirical MLR of the Hyades, we
require a measure of the apparent brightness of each star in the
visual band.  For this, we used the total $V$-band magnitude by
\cite{Joner:2006} ($V = 6.565 \pm 0.004$) and our spectroscopic light
ratios $\ell_{\rm A}/\ell_{\rm Ba}$ and $\ell_{\rm Bb}/\ell_{\rm Ba}$
from Section~\ref{sec:spectroscopy}, which correspond strictly to a
mean wavelength of 5187~\AA. To transform them to the $V$ band we
appealed to synthetic spectra based on PHOENIX model atmospheres from
the library of \cite{Husser:2013} for the same temperatures, surface
gravities, and metallicity adopted in our cross-correlation analysis.
For each pair of stars, the predicted flux ratio as a function of
wavelength was rescaled by adjusting the unknown ratio of the radii
until we reproduced the light ratio at 5187~\AA. We then integrated
over the $V$ band to obtain $(\ell_{\rm A}/\ell_{\rm Ba})_V = 2.00 \pm
0.15$ and $(\ell_{\rm Bb}/\ell_{\rm Ba})_V = 0.072 \pm 0.015$. With
the orbital parallax from Table~\ref{tab:deriv}, the absolute visual
magnitudes are then $M_V{\rm (A)} = 3.717 \pm 0.032$, $M_V{\rm (Ba)} =
4.470 \pm 0.055$, and $M_V{\rm (Bb)} = 7.33 \pm 0.24$.

\begin{figure}
\epsscale{1.15}
\plotone{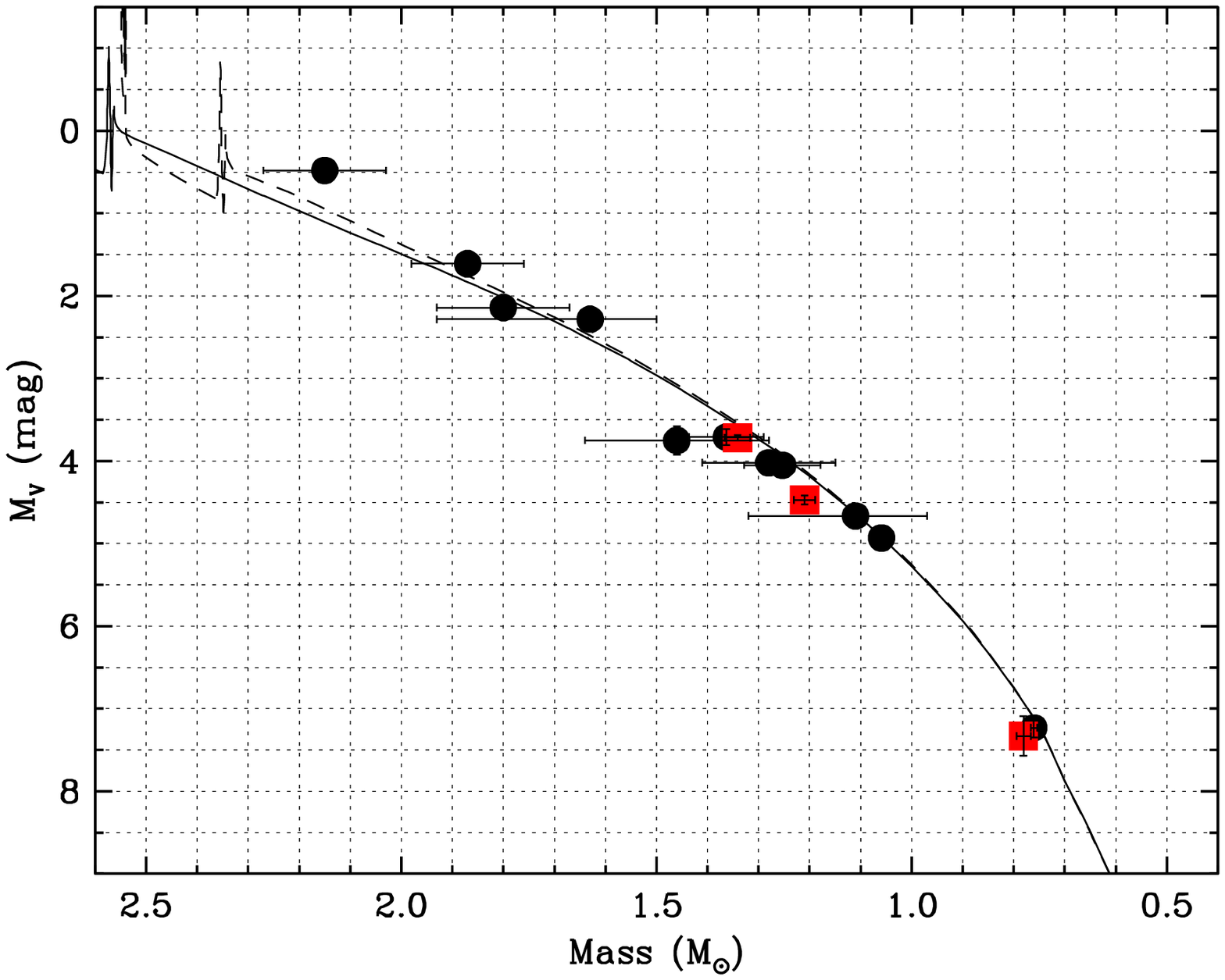}

\figcaption{Empirical mass-luminosity relation in the Hyades. The
  stars in \vstar\ are shown with red squares. Measurements for the
  other six binary systems in the cluster with dynamical mass
  determinations (V818\thinspace Tau, 51\thinspace Tau, 70\thinspace
  Tau, $\theta^1$\thinspace Tau, $\theta^2$\thinspace Tau, and
  80\thinspace Tau) are taken from Table~3 of \cite{Torres:2019}. The
  curves represent isochrones from the PARSEC series \citep{Chen:2014}
  for the known metallicity of the Hyades \citep[${\rm [Fe/H]} =
    +0.18$;][]{Dutra-Ferreira:2016} and ages of 625~Myr (solid line)
  and 800~Myr (dashed).\label{fig:MLR}}
\end{figure}

The mass and brightness of each \vstar\ component are shown in
Figure~\ref{fig:MLR} along with all other such measurements in the
Hyades taken from Table~3 of \cite{Torres:2019}. The 14 individual
mass estimates are broadly consistent with the stellar evolution
models shown by the solid and dashed curves. These models correspond,
respectively, to 625~Myr and 800~Myr isochrones from the PARSEC series
by \cite{Chen:2014}, calculated for the metallicity of the
Hyades. These models feature a modified temperature-opacity
  relation designed to provide a better fit to the mass-radius diagram
  of low-mass stars such as component Bb of \vstar, and also yield
  improved fits to the lower main-sequence in the color-magnitude
  diagrams of clusters.  We find that the new measurements for
\vstar\ fall slightly below the models, though it is difficult to say
whether this is because the masses are overestimated, the luminosities
underestimated (perhaps from a bias in the orbital parallax in the
upward direction), whether the models are too bright possibly due to
missing opacities, or some combination thereof.

%%%%%%%%%%%%%%%%%%%%%%%%%%%%%%%%%%%%%%%%%%%%%%%%%%%%%%%%%%%%%%%%%%%%%%%%
\section{Concluding remarks}
\label{sec:remarks}
%%%%%%%%%%%%%%%%%%%%%%%%%%%%%%%%%%%%%%%%%%%%%%%%%%%%%%%%%%%%%%%%%%%%%%%%

The detection of the lines of star Bb in our spectra has enabled the
determination of dynamical masses for all three members of the triple
system \vstar\ for the first time. Without that information, it would
only be possible to measure the mass of the primary star. The
combination of our extensive spectroscopic monitoring and existing
astrometric observations (visual and lunar occultation measurements,
and p.m.\ and parallax measures from \hip\ and \gaia) have enabled
mass estimates with formal precisions better than 2\% for the three
stars. These are the most precise mass determinations for any Hyades
binary with the exception of the eclipsing system V818\thinspace Tau.
The K dwarf tertiary component of \vstar\ happens to be nearly
identical in mass to the secondary of V818\thinspace Tau, and we find,
reassuringly, that so is its brightness.

Dynamical mass determinations in the Hyades now populate the empirical
MLR from about 0.76$~M_{\sun}$ to 2.1$~M_{\sun}$, though with a gap
between 0.78$~M_{\sun}$ (\vstar\ Bb) and 1.11$~M_{\sun}$ (80\thinspace
Tau B), where the models show the largest change in slope (see
Figure~\ref{fig:MLR}). Work is underway to identify additional cluster
members amenable to mass determinations that may help fill in this
gap.

\begin{acknowledgements}

Many of the spectroscopic observations of \vstar\ used here were
obtained with the assistance of M.\ Calkins, J.\ Caruso, P.\ Berlind,
G.\ Esquerdo, E.\ Horine, R.\ Mathieu, J.\ Peters, and J.\ Zajac. We
thank them all.  We are also grateful to Brian Mason for providing a
listing of the measurements of \vstar\ from the Washington Double Star
Catalog. The referee is thanked as well for the prompt review of
  our manuscript and helpful comments. This research has made use of
the SIMBAD and VizieR databases, operated at the CDS, Strasbourg,
France, of NASA's Astrophysics Data System Abstract Service, and of
the Washington Double Star Catalog maintained at the U.S.\ Naval
Observatory. The work has also made use of data from the European
Space Agency (ESA) mission
\gaia\ (\url{https://www.cosmos.esa.int/gaia}), processed by the
\gaia\ Data Processing and Analysis Consortium (DPAC,
\url{https://www.cosmos.esa.int/web/gaia/dpac/consortium}). Funding
for the DPAC has been provided by national institutions, in particular
the institutions participating in the \gaia\ Multilateral
Agreement. The computational resources used for this research include
the Smithsonian Institution's ``Hydra'' High Performance Cluster.

\end{acknowledgements}

\end{document}